\documentstyle[12pt,epsfig]{article}
\begin{document}
%\tighten

\thispagestyle{empty}
\begin{flushright}
SLAC-PUB-7613\\
hep-ph/9707545\\
August 1997
\end{flushright}
\vspace*{1.5cm}
\centerline{\Large\bf CP Violation in K and B Decays
\footnote{
Work supported by the Department of Energy under contract
DE-AC03-76SF00515.}}
\vspace*{1.5cm}
\centerline{{\sc Gerhard Buchalla}}
\bigskip
\centerline{\sl Stanford Linear Accelerator Center}
\centerline{\sl Stanford University, Stanford, CA 94309, U.S.A.}

\vspace*{1.2cm}
\centerline{\bf Abstract}
\vspace*{0.2cm}
\noindent 
We review basic aspects of the phenomenology of CP violation in
the decays of $K$ and $B$ mesons. In particular we discuss the
commonly used classification of CP violation -- CP violation
in the mass matrix, in the interference of mixing with decay, and
in the decay amplitude itself -- and the related notions of direct
and indirect CP violation. These concepts are illustrated with
explicit examples. We also emphasize the highlights of this field
including the clean observables $B(K_L\to\pi^0\nu\bar\nu)$ and
${\cal A}_{CP}(B\to J/\Psi K_S)$. The latter quantity serves to
demonstrate the general features of large, mixing induced CP
violation in $B$ decays.

\vspace*{2cm}
\centerline{\it Invited Talk presented at the}
\centerline{\it 6th Conference on the Intersections of Particle 
and Nuclear Physics} 
\centerline{\it Big Sky, Montana, May 27 -- June 2, 1997}

\vfill
\newpage
\pagenumbering{arabic}

\parindent=0pt

\section{Introduction}

Until today CP violation has only been observed in a few decay modes
of the long-lived neutral kaon, where it appears as a very small
(${\cal O}(10^{-3})$) effect. Despite continuing efforts since the
first observation of this phenomenon in 1964 and respectable progress
in both experiment and theory, our understanding of CP violation has
so far remained rather limited. Upcoming new experiments with $K$ and
$B$ mesons are likely to improve this situation substantially.
The great effort being invested into these studies is motivated by the
fundamental implications that CP violation has for our understanding of
nature: CP violation defines an absolute, physical distinction between
matter and antimatter. It is also one of the necessary conditions for the
dynamical generation of the observed baryon asymmetry in the universe.
In addition CP violation provides a testing ground for Standard Model
flavor dynamics -- the physics of quark masses and mixing.

The source of CP violation in the Standard Model (SM) is the
Cabibbo-Kobayashi-Maskawa (CKM) matrix $V$ entering the charged-current
weak interaction Lagrangian
\begin{equation}\label{gbuc:lcc}
{\cal L}_{CC}=\frac{g_W}{2\sqrt{2}} V_{ij} \bar u_i\gamma^\mu
(1-\gamma_5) d_j W^+_\mu + h.c.
\end{equation}
where $(u_1, u_2, u_3)\equiv (u,c,t)$, 
$(d_1, d_2, d_3)\equiv (d,s,b)$ are the mass eigenstates of the
six quark flavors and a summation over $i,j=1,2,3$ is understood.
The unitary CKM matrix ($V^\dagger V=1$) arises from diagonalizing
the quark mass matrix and relating the original weak eigenstates
of quark flavor to the physical mass eigenstates. The off-diagonal
elements of $V$ describe the strength of weak, charged current 
transitions between different generations of quarks.

In general, a $n\times n$ unitary matrix has $n^2$ free (real)
parameters. Not all of them are physical quantities in the present case
since one has the freedom of redefining the $2n$ fields $u_i$ and $d_j$
($i,j=1,\ldots, n$) by arbitrary phases $\alpha_i$ and $\beta_j$,
respectively. {}From (\ref{gbuc:lcc}) one sees that only the differences
$\alpha_i-\beta_j$ can affect $V$ in this redefinition. There are
$2n-1$ independent $\alpha_i-\beta_j$. The number of independent,
physical parameters that characterize $V$ is therefore
$n^2-(2n-1)=(n-1)^2$. Out of these $(n-1)^2$, $n(n-1)/2$, the number
of parameters of a real, orthogonal $n\times n$ matrix, represent
rotation angles. The remaining $(n-1)(n-2)/2$ are complex phases.
Obviously, then, for one or two generations of quarks the matrix $V$
can be chosen to be real. For the realistic case of three generations,
however, a physical complex phase is in general present in $V$
\cite{gbuc:KM}.
As a consequence, if this phase $\delta\not= 0,\pi$, the weak
interaction Lagrangian (\ref{gbuc:lcc}) is not invariant under CP.
(A further requirement for this to be true is that all three up-type
quark masses must be different from each other and the same must hold
for the down-type quarks. Otherwise an arbitrary unitary rotation
may be performed on the degenerate quark fields and the complex phase be 
removed. Also, none of the rotation angles must be $0$ or $\pi/2$.)
In the sections following this Introduction we will discuss how this
violation of CP symmetry at the level of the fundamental Lagrangian
manifests itself in observable CP asymmetries occuring in the weak
decays of $K$ and $B$ mesons. 

The CKM matrix has the following explicit form
\begin{equation}\label{gbuc:ckm}
V= \left( \begin{array}{ccc}
           V_{ud} & V_{us} & V_{ub} \\ 
           V_{cd} & V_{cs} & V_{cb} \\ 
           V_{td} & V_{ts} & V_{tb} 
         \end{array} \right) 
\simeq
   \left( \begin{array}{ccc}
           1-\lambda^2/2 & \lambda & A\lambda^3(\varrho- i \eta) \\ 
           -\lambda & 1-\lambda^2/2 & A \lambda^2 \\ 
           A\lambda^3(1-\varrho-i \eta) & -A\lambda^2 & 1
         \end{array} \right) 
\end{equation} 
where the second expression is a convenient parametrization in
terms of $\lambda$, $A$, $\varrho$ and $\eta$ due to Wolfenstein.
It is organized as a series expansion in powers of $\lambda=0.22$
(the sine of the Cabibbo angle) to exhibit the hierarchy among
the transitions between generations. Ordering transitions $i\to j$
according to decreasing strength, this hierarchy reads
$i\to i$ $>$ $1\to 2$ $>$ $2\to 3$ $>$ $1\to 3$, as is manifest in 
(\ref{gbuc:ckm}). The explicit parametrization shown in (\ref{gbuc:ckm})
is valid through order ${\cal O}(\lambda^3)$, an approxiamtion that is
sufficient for most practical applications. Higher order terms can
be taken into account if necessary \cite{gbuc:BBL}.

\begin{figure}[b!] % fig 1
 \vspace{5.5cm}
\includegraphics{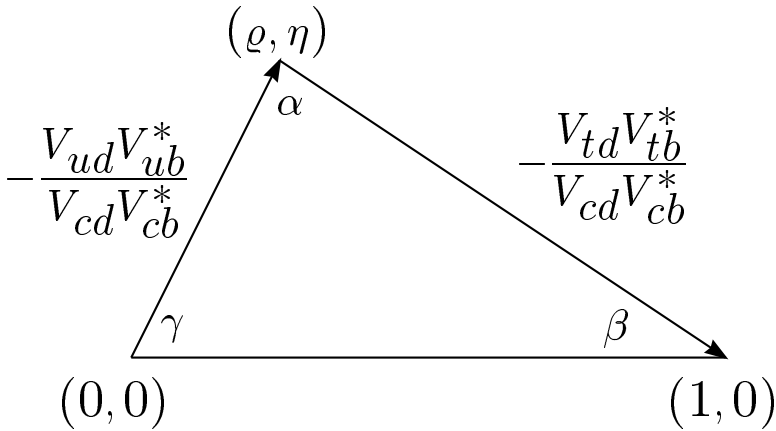}
%\centerline{\epsfig{file=gbfig1.eps,height=2.5in,width=3.5in}}
\vspace{10pt}
\caption{The normalized unitarity triangle in the $(\varrho,\eta)$
plane}
\label{gbuc:f:ut}
\end{figure}

The unitarity structure of the CKM matrix is conventionally displayed
in the so-called unitarity triangle (Fig. \ref{gbuc:f:ut}).
This triangle is a graphical representation of the unitarity relation
$V_{ud}V^*_{ub}+V_{cd}V^*_{cb}+V_{td}V^*_{tb}=0$
(normalized by $-V_{cd}V^*_{cb}$) in the complex plane of Wolfenstein
parameters $(\varrho,\eta)$.
The angles $\alpha$, $\beta$ and $\gamma$ of the unitarity triangle
are phase convention independent and can be determined in CP violation
experiments. The area of the unitarity triangle, which is proportional
to $\eta$, is a measure of CP nonconservation in the Standard Model.

The framework for a theoretical treatment of weak decays in general,
and CP violating processes in particular, is provided by low energy
effective Hamiltonians, which have the generic form
\begin{equation}\label{gbuc:heff}
{\cal H}_{eff}=\frac{G_F}{\sqrt{2}} V_{CKM}\sum_i
C_i(m_t,M_W/\mu,\alpha_s) Q_i
\end{equation}
Here $G_F$ is the Fermi constant, $V_{CKM}$ the appropriate combination
of CKM elements, $C_i$ are Wilson coefficients, which include also
strong interaction effects, and the $Q_i$ are local four-fermion
operators. (\ref{gbuc:heff}) provides a systematic approximation that
applies to processes where the relevant energy scale is much smaller than
the $W$-boson or the top quark mass, such as for instance $K$ and $B$
meson decays. An example for a typical operator is
$Q_2=(\bar su)_{V-A}(\bar ud)_{V-A}$, which appears in the analysis
of nonleptonic kaon decays. In essence the operators $Q_i$ are nothing
else than (effective) interaction vertices and the coefficients $C_i$
the corresponding coupling constants. The Hamiltonians (\ref{gbuc:heff})
can be derived from the fundamental Standard Model Lagrangian using
operator product expansion and renormalization group techniques.
They may be viewed as the modern generalization of the original
Fermi-theory of weak interactions. To calculate decay amplitudes,
matrix elements of the operators have to be evaluated between the
initial and final states under consideration. This is a problem
that involves nonperturbative QCD dynamics -- in general a difficult 
task not yet satisfactorily solved in many cases. The coefficients $C_i$
on the other hand are calculable in perturbation theory, as they 
incorporate the short distance contributions to the decay amplitude.
The factorization of short distance and
long distance contributions (Wilson coefficients and operator matrix 
elements, respectively), inherent in the effective Hamiltonian approach,
is a key feature of this framework.
Although we will not further elaborate on these issues here, the
effective Hamiltonian picture should be kept in mind as the theoretical
basis for weak decay phenomenology. A review of the current status of
this subject as well as an introduction to the basic concepts may be
found in \cite{gbuc:BBL}. For a general introduction to CP violation
see \cite{gbuc:CJ}.

The outline of this talk is as follows. After this Introduction we
briefly recall the physics of particle-antiparticle mixing, which is
crucial for the discussion of CP violation in neutral $K$ and $B$ meson
decays. We then describe a classification of CP violating phenomena in
$K$ and $B$ decays. To illustrate the concepts we will here use
kaon processes as specific examples. Subsequently we discuss the
rare decay mode $K_L\to\pi^0\nu\bar\nu$ and some of the basic issues
of CP violating asymmetries in $B$ decays. A short summary concludes
our presentation.

\section{Particle-Antiparticle Mixing}

\begin{figure}[b!] % fig 2
 \vspace{4.5cm}
\includegraphics{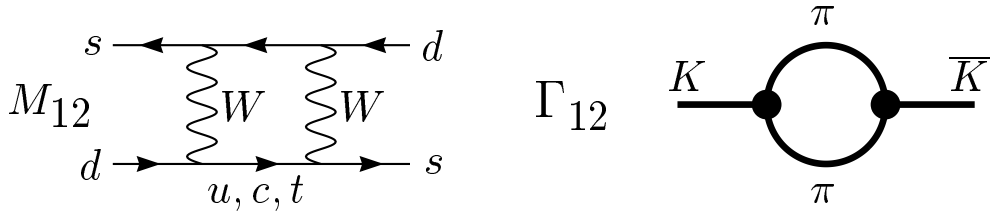}
%\centerline{\epsfig{file=gbfig2.eps,height=1.5in,width=3.5in}}
\vspace{10pt}
\caption{Diagrams contributing to $M_{12}$ and $\Gamma_{12}$ in
the neutral kaon system.}
\label{gbuc:f:mg12}
\end{figure}

Neutral $K$ and $B$ mesons can mix with their antiparticles through
second order weak interactions. They form two-state systems
($K^0-\bar K^0$, $B_d-\bar B_d$, $B_s-\bar B_s$) that are described
by Hamiltonian matrices $\hat H$ of the form
\begin{equation}\label{gbuc:hmg}
\hat H= \left( \begin{array}{cc}
           M_{11} & M_{12}  \\  
           M^*_{12} & M_{11} 
         \end{array} \right) 
-\frac{i}{2} \left( \begin{array}{cc}
           \Gamma_{11} & \Gamma_{12}  \\  
           \Gamma^*_{12} & \Gamma_{11} 
         \end{array} \right) 
\end{equation}
where CPT invariance has been assumed. The absorptive part
$\Gamma_{ij}$ of $\hat H$ accounts for the weak decay of
the neutral meson $F=K$, $B_d$, $B_s$. In Fig. \ref{gbuc:f:mg12}
we show typical diagrams that give rise to the off-diagonal elements
of $\hat H$ for the example of the kaon system.
Diagonalizing the Hamiltonian $\hat H$ yields the physical
eigenstates $F_{H,L}$. They are linear combinations of the strong
interaction eigenstates $F$ and $\bar F$ and can be written as
\begin{equation}\label{gbuc:fh}
F_H={\cal N}_{\bar\varepsilon}\left[(1+\bar\varepsilon)F +
  (1-\bar\varepsilon)\bar F\right]\equiv p F+ q \bar F
\end{equation}
\begin{equation}\label{gbuc:fl}
F_L={\cal N}_{\bar\varepsilon}\left[(1+\bar\varepsilon)F -
  (1-\bar\varepsilon)\bar F\right]\equiv p F - q \bar F
\end{equation}
with the normalization factor 
${\cal N}_{\bar\varepsilon}=1/\sqrt{2(1+|\bar\varepsilon|^2)}$.
Here $\bar\varepsilon$ is determined by
\begin{equation}\label{gbuc:epsmg}
\frac{1-\bar\varepsilon}{1+\bar\varepsilon}\equiv\frac{q}{p}=
\frac{M^*_{12}-\frac{i}{2}\Gamma^*_{12}}{\left(\Delta M
 +\frac{i}{2}\Delta\Gamma\right)/2}
\end{equation}
where $\Delta M$ and $\Delta\Gamma$ are the differences of the
eigenvalues $M_{H,L}-i\Gamma_{H,L}/2$ corresponding to the
eigenstates $F_{H,L}$
\begin{equation}\label{dmdg}
\Delta M\equiv M_H-M_L > 0 \qquad\qquad 
\Delta\Gamma\equiv \Gamma_L-\Gamma_H
\end{equation}
The labels $H$ and $L$ denote, respectively, the heavier and the lighter
eigenstate so that $\Delta M$ is positive by definition.
We employ here the CP phase convention $CP\cdot F=-\bar F$.
Using the SM results for $M_{12}$, $\Gamma_{12}$ and standard
phase conventions  for the CKM matrix (see (\ref{gbuc:ckm})), one
finds in the limit of CP conservation ($\eta=0$) that
$\bar\varepsilon=0$. With (\ref{gbuc:fh}), (\ref{gbuc:fl}) it follows
that $F_H$ is CP odd and $F_L$ is CP even in this limit, which is close
to realistic since CP violation is a small effect.
As we shall see explicitly later on, the real part of $\bar\varepsilon$
is a physical observable, while the imaginary part is not. In particular
$(1-\bar\varepsilon)/(1+\bar\varepsilon)$ is a phase convention dependent,
unphysical quantity.

Important characteristics of the three cases $F=K^0$, $B_d$, $B_s$
are collected in Table \ref{gbuc:t:kbb}.
\begin{table}
\centering
\caption{Important properties of neutral $K$ and $B$ meson systems.
Here $\Gamma\equiv(\Gamma_H+\Gamma_L)/2$. The kaon entries and
$\Delta M$ for $B_d$ are experimental results, the remaining numbers
theoretical expectations.}
\label{gbuc:t:kbb}
\begin{tabular}{|l|c|c|c|}
\hline
 & $K^0$ & $B_d$ & $B_s$ \\
\hline
$\Delta\Gamma/\Gamma$ & $2.0$, $\Gamma_L=579\cdot\Gamma_H$ &
    $\sim 0$ & $\sim 0.16\pm 0.10$  \\
\hline
$\Delta M/\Gamma$ & $0.95$ & $0.73\pm 0.05$ & $\sim 25\pm 15$ \\
\hline
\end{tabular}
\end{table}
A crucial feature of the kaon system is the very large difference in 
decay rates between the two eigenstates, the lighter eigenstate
decaying much more rapidly than the heavier one.
For the kaon system the states $F_L$ and $F_H$ are therefore
commonly denoted as short-lived ($K_S$) and long-lived ($K_L$)
eigenstates, respectively. The same hierarchy in decay rates is
expected for the $B_s$ mesons, although far less pronounced
as $\Gamma_H/\Gamma_L={\cal O}(1)$. In the case of $B_d$ 
$\Delta\Gamma/\Gamma$ is essentially negligible. The labeling of 
eigenstates as heavy/light is therefore more common for $B$ mesons.
The basic reason for this pattern is the small number of decay channels 
for the neutral kaons. Decay into the predominant
CP even two-pion final states $\pi^+\pi^-$, $\pi^0\pi^0$ is only
available for $K_S$, but not (to first approximation) for the
(almost) CP odd state $K_L$. The latter can decay into three pions,
which however is kinematically strongly suppressed, leading to a much
longer $K_L$ lifetime. This somewhat accidental feature is absent for
$B$ mesons, which have many more decay modes due to their larger mass.
We may summarize this discussion by noting that in general the following
correspondence holds for the eigenstates of the neutral $K$ and $B$
systems. One has {\bf H}eavy={\bf L}ong-lived$\approx$CP odd, and
{\bf L}ight={\bf S}hort-lived$\approx$CP even, where the CP assignments
are only approximate due to CP violation.

\section{Classification of CP Violation}

The CP noninvariance of the fundamental weak interaction Lagrangian
leads to a violation of CP symmetry at the phenomenological level, in
particular in decays of $K$ and $B$ mesons. For instance, processes
forbidden by CP symmetry may occur or transitions related to each
other by CP conjugation may have a different rate.
The phenomenology of CP violating decays is very rich, already for
kaons and even more so for $B$ mesons. In this situation it is
certainly helpful to have a classification of the various possible
mechanisms at hand. One that is commonly used in the literature on
this subject employs the following terminology.

a) {\it CP violation in the mixing matrix.\/}
This type of effect is based on CP violation in the two-state
mixing Hamiltonian $\hat H$ (\ref{gbuc:hmg}) itself and is measured by the
observable quantity ${\rm Im}(\Gamma_{12}/M_{12})$.
It is related to a change in flavor by two units,
$\Delta S (\Delta B)=2$.

b) {\it CP violation in the decay amplitude.\/} This class of phenomena
is characterized by CP violation originating directly in the amplitude
for a given decay. It is entirely independent of particle-antiparticle
mixing and can therefore occur for charged mesons ($K^\pm$, $B^\pm$)
as well. Here the transitions have $\Delta S (\Delta B)=1$. 

c) {\it CP violation in the interference of mixing and decay.\/}
In this case the interference of two amplitudes, necessary in general
to induce observable CP violation, takes place between the mixing amplitude
and the decay amplitude in decays of neutral $K$ and $B$ mesons.
This very important class is sometimes also refered to as
{\it mixing-induced\/} CP violation, a terminology not to be confused
with a).

Complementary to this classification is the widely used notion of
{\it direct\/} versus {\it indirect\/} CP violation. It is motivated
historically by the hypothesis of a new superweak interaction
\cite{gbuc:WO1,gbuc:WO2}, that was proposed as early as 1964
by Wolfenstein to account for the CP violation observed in 
$K_L\to\pi^+\pi^-$ decay. This new CP violating interaction would lead
to a local four-quark vertex that changes flavor quantum number 
(strangeness or beauty) by two units. Its only effect would be a CP
violating contribution to $M_{12}$, so that all observed CP violation
could be attributed to particle-antiparticle mixing alone.
Today, after the advent of the three generation SM, the CKM mechanism
of CP violation appears more natural. In principal the superweak
scenario represents a logical possibility, leading to a different pattern
of observable CP violation effects. In fact, all experimental
measurements available to date are still consistent with the superweak
hypothesis.
\\
Now, any CP violating effect that can be entirely assigned to CP
violation in $M_{12}$ (as for the superweak case) is termed
{\it indirect CP violation\/}. Conversely, any effect that can not be 
described in this way and explicitly requires CP violating phases in
the decay amplitude itself is called {\it direct CP violation\/}.
It follows that class a) represents indirect, class b) direct CP
violation. Class c) contains aspects of both. In this latter case
the magnitude of CP violation observed in any one decay mode (within the
neutral kaon system, say) could by itself be ascribed to mixing, thus
corresponding to an indirect effect. On the other hand, a difference in
the degree of CP violation between two different modes would reveal
a direct effect.

The classification a) -- c) is especially common in the context of
$B$ physics but it applies to kaon physics as well. To emphasize this
point and to provide concrete examples for the above general concepts,
we will next illustrate these classes by important applications in
kaon decays. We will also use this opportunity to discuss several aspects
of kaon CP violation in more detail. After all $K$ physics is the area
from which our entire present experimental knowledge of CP violation
derives. For a general review of CP violation in kaon decays
see \cite{gbuc:DI}.

\subsection*{a) -- Lepton Charge Asymmetry}

The lepton charge asymmetry in semileptonic $K_L$ decay is an 
example for CP violation in the mixing matrix. It is probably
the most obvious manifestation of CP nonconservation in kaon decays.
The observable considered here reads ($l=e$ or $\mu$)
\begin{eqnarray}\label{gbuc:lca}
\Delta &=& \frac{\Gamma(K_L\to\pi^-l^+\nu)-\Gamma(K_L\to\pi^+l^-\bar\nu)}{
  \Gamma(K_L\to\pi^-l^+\nu)+\Gamma(K_L\to\pi^+l^-\bar\nu)}=
\frac{|1+\bar\varepsilon|^2-|1-\bar\varepsilon|^2}{
   |1+\bar\varepsilon|^2+|1-\bar\varepsilon|^2} \nonumber \\
&\approx& 2{\rm Re}\ \bar\varepsilon
\approx\frac{1}{4}{\rm Im}\frac{\Gamma_{12}}{M_{12}}
\end{eqnarray}
If CP was a good symmetry of nature, $K_L$ would be a CP eigenstate and
the two processes compared in (\ref{gbuc:lca}) were related by a
CP transformation. The rate difference $\Delta$ should vanish.
Experimentally one finds however \cite{gbuc:PDG}
\begin{equation}\label{gbuc:dexp}
\Delta_{exp}=(3.27\pm 0.12)\cdot 10^{-3}
\end{equation}
a clear signal of CP violation. The second equality in (\ref{gbuc:lca})
follows from (\ref{gbuc:fh}), as applied to $K_L$, noting that the
positive lepton $l^+$ can only originate from $K\sim(\bar sd)$,
$l^-$ only from $\bar K\sim (\bar ds)$. This is true to leading order
in SM weak interactions and holds to sufficient accuracy for our purpose.
The charge of the lepton essentially serves to tag the strangeness
of the $K$, thus picking out either only the $K$ or only the $\bar K$
component. Any phase in the semileptonic amplitudes is irrelevant
and the CP violation effect is purely in the mixing matrix itself.
In fact, as indicated in (\ref{gbuc:lca}), $\Delta$ is determined by
${\rm Im}(\Gamma_{12}/M_{12})$, the physical measure of CP
violation in the mixing matrix.
\\
{}From (\ref{gbuc:dexp}) we see that $\Delta>0$. This empirical fact
can be used to define positive electric charge in an absolute, physical
sense. Positive charge is the charge of the lepton more copiously
produced in semileptonic $K_L$ decay. This definition is unambiguous
and would even hold in an antimatter world. Also, using 
some parity violation experiment, this result implies in addition an
absolute definition of left and right. These are quite remarkable
facts. They clearly provide part of the motivation to try to learn
more about the origin of CP violation.

\subsection*{b) -- CP Violation in the Decay Amplitude}

Observable CP violation may also occur through interference effects 
in the decay amplitudes themselves (pure direct CP violation).
This case is conceptually perhaps the simplest mechanism for
CP violation and the basic features are here particularly transparent.
Consider a situation where two different components contribute to the
amplitude of a $K$ meson decaying into a final state $f$
\begin{equation}\label{gbuc:akf}
A\equiv A(K\to f)=A_1 e^{i\delta_1}e^{i\phi_1}+
                  A_2 e^{i\delta_2}e^{i\phi_2}
\end{equation}
Here $A_i$ ($i=1,2$) are real amplitudes and $\delta_i$ are complex
phases from CP conserving interactions. The $\delta_i$ are usually
strong interaction rescattering phases. Finally the $\phi_i$ are weak
phases, that is phases coming from the CKM matrix in the SM.
The corresponding amplitude for the CP conjugated process
$\bar K\to\bar f$ then reads (the explicit minus signs are due to
our convention $CP\cdot K=-\bar K$, ($CP\cdot f=\bar f$))
\begin{equation}\label{gbuc:akfb}
\bar A\equiv A(\bar K\to \bar f)=-A_1 e^{i\delta_1}e^{-i\phi_1}-
                  A_2 e^{i\delta_2}e^{-i\phi_2}
\end{equation}
Since now all quarks are replaced by antiquarks (and vice versa) compared
to (\ref{gbuc:akf}), the weak phases change sign. The CP invariant
strong phases remain the same.
From (\ref{gbuc:akf}) and (\ref{gbuc:akfb}) one finds immediately
\begin{equation}\label{gbuc:aab}
|A|^2-|\bar A|^2 \sim A_1 A_2 \sin(\delta_1-\delta_2)
  \sin(\phi_1-\phi_2)
\end{equation}
The conditions for a nonvanishing difference between the decay rates
of $K\to f$ and the CP conjugate $\bar K\to\bar f$, that is direct
CP violation, can be read off from (\ref{gbuc:aab}).
There need to be two interfering amplitudes $A_1$, $A_2$ and these
amplitudes must simultaneously have both different weak ($\phi_i$) and
different strong phases ($\delta_i$). Although the strong interaction
phases can of course not generate CP violation by themselves, they are
still a necessary requirement for the weak phase differences to show up
as observable CP asymmetries. It is obvious from (\ref{gbuc:akf}) and
(\ref{gbuc:akfb}) that in the absence of strong phases $A$ and $\bar A$
would have the same absolute value despite their different weak phases,
since then $A=-\bar A^*$.

\begin{figure}[b!] % fig 3
 \vspace{4.5cm}
\includegraphics{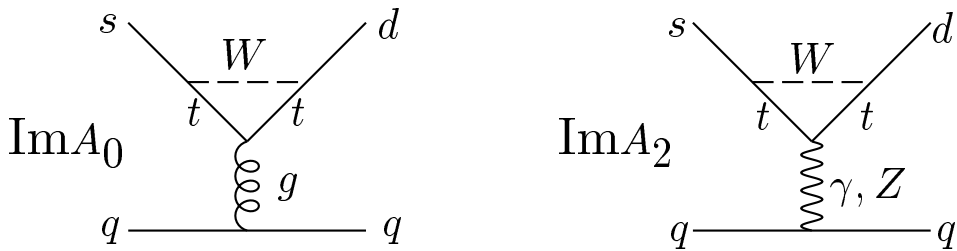}
%\centerline{\epsfig{file=gbfig3.eps,height=1.5in,width=3.5in}}
\vspace{10pt}
\caption{Gluon penguin and electroweak penguin diagram contributions
to the parameter $\varepsilon'$.}
\label{gbuc:f:gewp}
\end{figure}

A specific example is given by the decays $K(\bar K)\to\pi^+\pi^-$
(here $f=\pi^+\pi^-=\bar f$). The amplitudes can be written as
\begin{eqnarray}\label{gbuc:apm}
A_{+-} &=& \sqrt{\frac{2}{3}}A_0 e^{i\delta_0}+
  \frac{1}{\sqrt{3}}A_2 e^{i\delta_2} \nonumber \\
\bar A_{+-} &=& -\sqrt{\frac{2}{3}}A^*_0 e^{i\delta_0}-
  \frac{1}{\sqrt{3}}A^*_2 e^{i\delta_2} 
\end{eqnarray}
where $A_{0,2}=\langle\pi\pi(I=0,2)|{\cal H}_W|K\rangle$
are the transition amplitudes of $K$ to the isospin-0 and
isospin-2 components of the $\pi^+\pi^-$ final state. They still
include the weak phases, but the strong phases have been factored out and 
written explicitly in (\ref{gbuc:apm}). Taking the modulus squared of the
amplitudes we get
\begin{eqnarray}\label{gbuc:reep}
\frac{\Gamma(K\to\pi^+\pi^-)-\Gamma(\bar K\to\pi^+\pi^-)}{
  \Gamma(K\to\pi^+\pi^-)+\Gamma(\bar K\to\pi^+\pi^-)} &=&
\sqrt{2}\sin(\delta_0-\delta_2)\frac{{\rm Re}A_2}{{\rm Re}A_0}
\left(\frac{{\rm Im}A_2}{{\rm Re}A_2}-\frac{{\rm Im}A_0}{{\rm Re}A_0}
\right) \nonumber \\
&=& 2\ {\rm Re}\ \varepsilon'
\end{eqnarray}
The quantity so defined is just twice the real part of the famous 
parameter $\varepsilon'$, the measure of direct CP violation in
$K\to\pi\pi$ decays.
The real parts of $A_{0,2}$ can be extracted from experiment.
The imaginary parts have to be calculated using the effective
Hamiltonian formalism briefly sketched in the Introduction. Ultimately
the amplitudes derive from quark level diagrams. The most important
contributions, the gluon penguin and the electroweak penguin, are
depicted in Fig. \ref{gbuc:f:gewp}.
The importance of the electroweak penguin graph might be surprising
at first sight; after all it is a contribution suppressed by small
electroweak couplings compared to the strong interaction effect
represented by the gluon penguin diagram. However, there are several
circumstances that actually conspire so as to enhance the impact of
the electroweak sector substantially. First of all, the electroweak
diagrams contribute to ${\rm Im}A_2$, in contrast to the gluon penguins,
which correspond to pure $\Delta I=1/2$ operators (the gluon
coupling conserves isospin) and can only lead to an isospin-0 final state,
starting from a kaon with isospin $1/2$. Furthermore, the suppression
$\sim\alpha/\alpha_s$ from coupling constants is largely compensated
by the fact that ${\rm Re}A_0\gg {\rm Re}A_2$, reflecting the empirical
$\Delta I=1/2$ rule in nonleptonic kaon decays. In addition the electroweak
contribution grows strongly with the top quark mass 
\cite{gbuc:FR,gbuc:BBH} and turns out to be quite substantial for the
actual value $\bar m_t(m_t)=167\ GeV$ 
($\overline{MS}$-mass). Entering with sign opposite to the (positive)
gluon penguin contribution, the electroweak penguin contribution tends 
to cancel the latter. This feature makes a precise theoretical prediction
of $\varepsilon'$, which anyhow suffers from large hadronic uncertainties,
even more difficult.
The typical order of magnitude of $\varepsilon'$ can however be understood
from (\ref{gbuc:reep}). The size of ${\rm Im}A_i/{\rm Re}A_i$ is
essentially determined by the small CKM parameters that carry the complex 
phase and which are related to the top quark in the loop diagrams from
Fig. \ref{gbuc:f:gewp}. Roughly speaking
${\rm Im}A_i/{\rm Re}A_i\sim {\rm Im}V^*_{ts}V_{td}\sim 10^{-4}$.
Empirically we have, from the $\Delta I=1/2$ rule,
${\rm Re}A_2/{\rm Re}A_0\sim 10^{-2}$. This leads to a natural size
of $\varepsilon'$ of $\sim 10^{-6}$, or possibly even smaller due to the
cancellations mentioned before.

We should stress that the quantity in (\ref{gbuc:reep}) is not the
observable actually used to determine $\varepsilon'$ experimentally.
We have discussed it here because it is of conceptual interest as
the simplest manifestation of $\varepsilon'$. The realistic analysis
requires a more general consideration of $K_L, K_S\to\pi\pi$ decays
to which we will turn in the following paragraph.

\subsection*{c) -- Mixing Induced CP Violation in $K\to\pi\pi$:
  $\varepsilon$, $\varepsilon'$}

In this section we will illustrate the concept of mixing-induced
CP violation with the example of $K\to\pi\pi$ decays. These are
important processes, since CP violation has first been seen in
$K_L\to\pi^+\pi^-$ and as of today our most precise experimental
knowledge about this phenomenon still comes from the study of
$K\to\pi\pi$ transitions. There are two distinct final states and in a
strong interaction eigenbasis the transitions are 
$K^0,\bar K^0\to\pi\pi(I=0), \pi\pi(I=2)$, with definite isospin for
$\pi\pi$. Alternatively, using the physical eigenbasis for both
initial and final states, one has $K_L,K_S\to\pi^+\pi^-,\pi^0\pi^0$.

Consider next the amplitude for $K_L$ going into the CP even state
$\pi\pi(I=0)$, which can proceed via $K$ 
($\sim(1+\bar\varepsilon)A_0$) or via $\bar K$
($\sim(1-\bar\varepsilon)A^*_0$). Hence (to first order in small
quantities)
\begin{equation}\label{gbuc:ebe}
A(K_L\to\pi\pi(I=0))\sim 
(1+\bar\varepsilon)A_0 e^{i\delta_0}-
(1-\bar\varepsilon)A^*_0 e^{i\delta_0}\sim
\bar\varepsilon+i\frac{{\rm Im}A_0}{{\rm Re}A_0}=\varepsilon
\end{equation}
This defines the parameter $\varepsilon$, characterizing mixing-induced
CP violation. Note that $\varepsilon$ involves a component from
mixing ($\bar\varepsilon$) as well as from the decay amplitude
(${\rm Im}A_0/{\rm Re}A_0$). Neither of those is physical separately,
but $\varepsilon$ is. Note also that the physical quantity
${\rm Re}\bar\varepsilon$ discussed above satisfies
${\rm Re}\bar\varepsilon={\rm Re}\varepsilon$.
More generally one can form the following two CP violating observables
\begin{equation}\label{gbuc:epm0}
\eta_{+-}=\frac{A(K_L\to\pi^+\pi^-)}{A(K_S\to\pi^+\pi^-)}
\qquad\qquad
\eta_{00}=\frac{A(K_L\to\pi^0\pi^0)}{A(K_S\to\pi^0\pi^0)}
\end{equation}
These amplitude ratios involve the physical initial and final states
and are directly measurable in experiment. They are related to
$\varepsilon$ and $\varepsilon'$ through
\begin{equation}\label{gbuc:eteps}
\eta_{+-}=\varepsilon+\varepsilon'\qquad\qquad
\eta_{00}=\varepsilon -2\varepsilon'
\end{equation}
The phase of $\varepsilon$ is given by
$\varepsilon=|\varepsilon|\exp(i\pi/4)$. The relative phase between
$\varepsilon'$ and $\varepsilon$ can be determined theoretically.
It is close to zero so that to very good approximation
$\varepsilon'/\varepsilon={\rm Re}\varepsilon'/\varepsilon$.

Both $\eta_{+-}$ and $\eta_{00}$ measure mixing-induced CP violation
(interference between mixing and decay). Each of them considered 
separately could be attributed to CP violation in $K-\bar K$ mixing
and would therefore represent indirect CP violation. On the other hand,
a nonvanishing difference $\eta_{+-}-\eta_{00}=3\varepsilon'\not=0$
is a signal of direct CP violation.
Experimentally one has \cite{gbuc:PDG}
\begin{equation}\label{gbuc:eps}
|\varepsilon|=(2.282\pm 0.019)\cdot 10^{-3}
\end{equation}
Theoretically $\varepsilon$ is related to the first diagram shown
in Fig. \ref{gbuc:f:mg12}. Comparison of the theoretical expression
\cite{gbuc:BBL} with the experimental result yields an important
constraint on the CKM phase $\delta$ (this is the phase of the
CKM matrix in standard parametrization \cite{gbuc:PDG}; it
coincides with the phase $\gamma$ of the unitarity triangle).
The quantity $\varepsilon'$ can be measured as the ratio
${\rm Re}\varepsilon'/\varepsilon\approx\varepsilon'/\varepsilon$
using the double ratio of rates
\begin{equation}\label{repe}
\left|\frac{\eta_{+-}}{\eta_{00}}\right|^2\doteq
1+6\ {\rm Re}\frac{\varepsilon'}{\varepsilon}
\end{equation}
Currently the following measurements are available
\begin{equation}\label{gbuc:epex}
{\rm Re}\frac{\varepsilon'}{\varepsilon}=
\left\{ \begin{array}{ll}
        (23\pm 7)\cdot 10^{-4} & \mbox{CERN NA31} \\
        (7.4\pm 5.9)\cdot 10^{-4} & \mbox{FNAL E731}
        \end{array}
\right.
\end{equation}
These results are somewhat inconclusive and it remains presently still
open whether or not a direct CP violation effect exists in
$K\to\pi\pi$ decays. As mentioned before, the theoretical
predictions suffer from large hadronic uncertainties. A representative
range from a recent analysis of Buras et al. \cite{gbuc:BJL} is
\begin{equation}\label{gbuc:epe}
2\cdot 10^{-4}\leq \varepsilon'/\varepsilon \leq 19\cdot 10^{-4}
\end{equation}
Similar results have been obtained by other groups 
\cite{gbuc:CFMRS,gbuc:BEF2,gbuc:HEI}. 
Currently running or future experiments at CERN, FNAL and Frascati
aim at an improved sensitivity of 
$\Delta\varepsilon'/\varepsilon\approx 10^{-4}$. 
If $\varepsilon'/\varepsilon$
is not too small, the new round of
measurements has a good chance to finally resolve the question of
direct CP violation in $K\to\pi\pi$ experimentally.

\section{The Rare Decay $K_L\to\pi^0\nu\bar\nu$}

One of the most promising opportunities for future studies of
flavor physics and CP violation is the rare decay 
$K_L\to\pi^0\nu\bar\nu$.
This process combines strongly motivated phenomenological interest
(sensitivity to high energy scales, top quark mass and CKM couplings)
with a situation where all theoretical uncertainties are
exceedingly well under control. With these features 
$K_L\to\pi^0\nu\bar\nu$ is unparalleled in the phenomenology of weak 
decays.

$K_L\to\pi^0\nu\bar\nu$ is a flavor-changing neutral current process,
induced at one-loop order in the SM.
It proceeds entirely through short distance weak
interactions because the neutrinos can couple only to heavy gauge
bosons ($W$, $Z$). The transition can be effectively described by a
local $(\bar sd)_{V-A}(\bar\nu\nu)_{V-A}$ interaction (and $h.c.$),
whose coupling strength is calculable from the SM. This interaction is
semileptonic and the required hadronic matrix element
$\langle\pi^0|(\bar sd)_V|K^0\rangle$ can be extracted from the well
measured decay $K^+\to\pi^0e^+\nu$ using isospin symmetry.
The knowledge of short distance QCD effects at next-to-leading order
(${\cal O}(\alpha_s)$) \cite{gbuc:BB2}, essentially eliminates the
dominant theoretical uncertainty in this decay mode from scale
dependence. The process is theoretically under control to an accuracy
of better than $\pm 3\%$.

In the limit of conserved CP, the relevant hadronic matrix element
would be $\langle\pi^0|(\bar sd)_V+(\bar ds)_V|K_L\rangle$.
Because of the CP properties of $K_L$, $\pi^0$ and the transition
current this matrix element is zero in this limit. 
In the SM $K_L\to\pi^0\nu\bar\nu$ therefore measures a violation
of CP symmetry.
It belongs to the class of mixing-induced CP violation.
Considering the amplitude ratio
$\eta_{\pi^0\nu\bar\nu}=A(K_L\to\pi^0\nu\bar\nu)/A(K_S\to\pi^0\nu\bar\nu)$, 
which is analogous to $\eta_{+-}$ for $K\to\pi^+\pi^-$
(\ref{gbuc:epm0}), one finds 
$\eta_{\pi^0\nu\bar\nu}={\cal O}(1)$ in the SM, essentially because
$K\to\pi^0\nu\bar\nu$ is a rare decay. Thus we have
$\eta_{\pi^0\nu\bar\nu}\gg\eta_{+-}={\cal O}(10^{-3})$, which means
that $K_L\to\pi^0\nu\bar\nu$ is a signal of very large
direct CP violation within the SM.
The branching ratio $B(K_L\to\pi^0\nu\bar\nu)$ is proportional to
$({\rm Im}V^*_{ts}V_{td})^2$, which makes it an ideal measure of
${\rm Im}V^*_{ts}V_{td}$ or the parameter $\eta$.

The current SM prediction for the branching ratio is
$B(K_L\to\pi^0\nu\bar\nu)=(2.8\pm 1.7)\cdot 10^{-11}$ 
\cite{gbuc:BJLnew}, where the sizable range reflects our presently
still quite limited knowledge of CKM parameters, but not intrinsic
theoretical uncertainties, which are negligible.
Using the experimental limit on $K^+\to\pi^+\nu\bar\nu$,
a model independent upper bound can be set at
$B(K_L\to\pi^0\nu\bar\nu)< 1.1\cdot 10^{-8}$ \cite{gbuc:GN}.
Current experimental searches, not optimized for this process, have
yielded a (published \cite{gbuc:PDG}) upper bound of
$5.8\cdot 10^{-5}$ (Fermilab E799). Dedicated experiments will aim at
an actual measurement of $K_L\to\pi^0\nu\bar\nu$ in the future.
A proposal already exists at Brookhaven (BNL E926) and there are
further plans at Fermilab and KEK.

\section{CP Violation in B Decays}

Decays of $B$ mesons offer a wide range of possibilities to expand
our knowledge of CP violation and to test further what we have
learned from the kaon system. Among those are truly superb opportunities
with esssentially no theoretical uncertainty and predicted large
CP asymmetries. The prototype observable is the time-dependent
CP asymmetry in $B_d(\bar B_d)\to J/\Psi K_S$, which is without doubt
the highlight of this field. We will first focus on this case in the 
following because of its importance and because it exhibits the 
characteristic features of a large class of CP violating observables
in $B$ physics. We will briefly mention further possibilities later on.

\subsection{$B_d\to J/\Psi K_S$}

\begin{figure}[t!] % fig 4
 \vspace{4.0cm}
\includegraphics{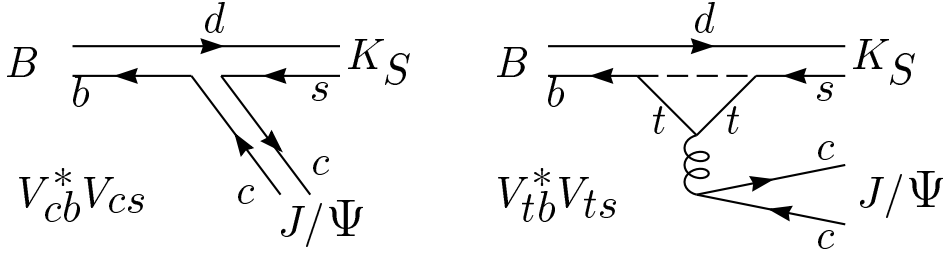}
%\centerline{\epsfig{file=gbfig4.eps,height=1.5in,width=3.5in}}
\vspace{10pt}
\caption{Representative diagrams contributing to $B_d\to J/\psi K_S$.}
\label{gbuc:f:trpg}
\end{figure}

\begin{figure}[b!] % fig 5
 \vspace{4.5cm}
\includegraphics{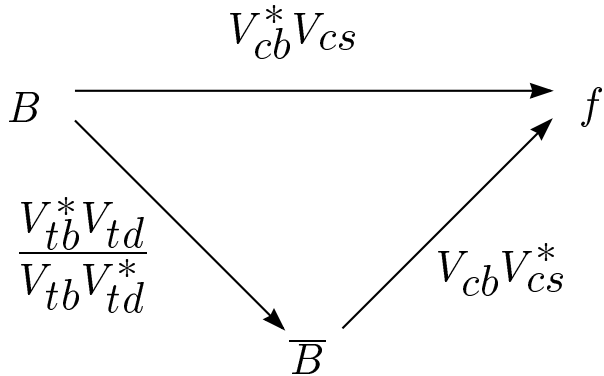}
%\centerline{\epsfig{file=gbfig5.eps,height=2.0in,width=3.5in}}
\vspace{10pt}
\caption{Possible decay paths for an initial $B$ meson to decay into
the final state $f=J/\Psi K_S$ that is common to $B$ and $\bar B$.}
\label{gbuc:f:bbf}
\end{figure}

The CP asymmetry in $B_d\to J/\Psi K_S$ belongs to the class of
mixing-induced CP violation, that is CP violation in the interference
of mixing and decay. In the kaon system an essentially pure beam of
a definite eigenstate, the $K_L$, can easily be produced due to the vast
difference in lifetimes between $K_L$ and $K_S$, which is ideal for
CP violation studies. Since the lifetime difference between eigenstates is
negligibly small for the $B_d-\bar B_d$ system, the same method can not
be applied in this case. Instead explicit flavor tagging
(determination of the flavor of one of the $B$ mesons (produced in pairs),
for instance by means of the lepton charge in the semileptonic decay of
the other) is required and one has to consider the time dependence of
$B-\bar B$ mixing.\footnote{This latter strategy can in principle also
be used for neutral kaons and is in fact the method realized in the
CPLEAR experiment at CERN (see M. Mikuz, these proceedings).}
Solving the time dependent Schr\"odinger equation
with the mixing Hamiltonian $\hat H$ (\ref{gbuc:hmg}), and neglecting
$\Delta\Gamma$, one has
\begin{equation}\label{gbuc:bt}
B(t)=e^{-iMt-\frac{1}{2}\Gamma t}
\left[\cos\frac{\Delta M t}{2}\ B-\frac{q}{p}i\sin\frac{\Delta M t}{2}\
\bar B\right]
\end{equation}
\begin{equation}\label{gbuc:btb}
\bar B(t)=e^{-iMt-\frac{1}{2}\Gamma t}
\left[\cos\frac{\Delta M t}{2}\ \bar B-\frac{p}{q}i\sin\frac{\Delta M t}{2}\
B\right]
\end{equation}
$B(t)$ and $\bar B(t)$ are the time evolved states that started out as
flavor eigenstates $B$ and $\bar B$, respectively, at time $t=0$.
\\
The CP asymmetry in $B_d\to J/\Psi K_S$ is the prime example of the
important class of asymmetries in neutral $B$ mesons decaying into a
CP eigenstate, in this case $f=J/\Psi K_S$, which is CP odd.
There are two basic contributions to the decay amplitude, distinguished
by the combination of CKM parameters $V^*_{cb}V_{cs}$ or
$V^*_{tb}V_{ts}$. Representative diagrams are shown in 
Fig. \ref{gbuc:f:trpg}. 
The third possible factor $V^*_{ub}V_{us}$ can be expressed in terms of
the above two by CKM unitarity, 
$V^*_{ub}V_{us}=-V^*_{cb}V_{cs}-V^*_{tb}V_{ts}$.
Choosing the latter two as independent parameters is useful in the 
present case, since $V^*_{ub}V_{us}$ is Cabibbo suppressed.
\\
A crucial feature of the $B_d\to J/\Psi K_S$ mode is that the relative
weak phase between $V^*_{cb}V_{cs}$ and $V^*_{tb}V_{ts}$ is negligibly
small. Consequently the $B_d\to J/\Psi K_S$ amplitude can to excellent
approximation be represented as
$A(B_d\to J/\Psi K_S)=V^*_{cb}V_{cs}\cdot A_{red}$ as far as the
weak phase structure is concerned. The quantity $A_{red}$ involves
nontrivial hadronic dynamics, but it will drop out when forming the ratio
that defines the asymmetry (see (\ref{gbuc:acp}) below). This fact lies
at the bottom of the theoretically clean nature of the
$B_d\to J/\Psi K_S$ asymmetry.
\\
Using this property of the amplitude we can now see how the mixing-induced
asymmetry comes about. As illustrated in Fig. \ref{gbuc:f:bbf},
an initial $B$ state can decay to the CP self-conjugate final state $f$
via two different paths: directly ($B\to f$), or through mixing
($B\to \bar B\to f$), since the same final state can be reached by both
$B$ and $\bar B$.
The mixing phase phase ($B\to\bar B$) is determined by the box diagram,
similar to the first graph in Fig. \ref{gbuc:f:mg12}, and reads
$(V^*_{tb}V_{td})^2/|V^*_{tb}V_{td}|^2\equiv 
V^*_{tb}V_{td}/(V_{tb}V^*_{td})$. The two different decay paths therefore
have a relative phase of
$V^*_{tb}V_{td}V_{cb}V^*_{cs}/(V_{tb}V^*_{td}V^*_{cb}V_{cs})=
\exp(-2i\beta)$.
The CP conjugate situation (starting out with $\bar B$) has the opposite
phase. Putting everything together (using (\ref{gbuc:bt}),
(\ref{gbuc:btb})) one finds the time-dependent asymmetry
\begin{eqnarray}\label{gbuc:acp}
{\cal A}_{CP}(B_d\to J/\Psi K_S) &\equiv&
\frac{\Gamma(B(t)\to\Psi K_S)-\Gamma(\bar B(t)\to\Psi K_S)}{
  \Gamma(B(t)\to\Psi K_S)+\Gamma(\bar B(t)\to\Psi K_S)}\nonumber \\
 &=& -\sin 2\beta\cdot\sin\Delta M t
\end{eqnarray}
A few points about this result are worth emphasizing, the first two
of which summarize the basic reasons why
${\cal A}_{CP}(B_d\to J/\Psi K_S)$ plays such an important role in
flavor physics.
\begin{itemize}
\item
As mentioned before, the part of the amplitude containing the dependence
on the uncalculable hadronic dynamics has canceled out in the 
asymmetry. The asymmetry depends only on the CKM quantity $\sin 2\beta$.
This result holds to within a theoretical uncertainty of less than $1\%$.
\item
The effect is quite large in the SM, where one expects approximately
$\sin 2\beta\approx 0.6\pm 0.2$. This information comes from the
observed CP violation in the kaon system, which implies that the
CP phase $\eta$ must not be too small. It follows
(see Fig. \ref{gbuc:f:ut}) that also $\sin 2\beta$ has to be sizable.
\item
Two a priori unrelated features of the fundamental SM parameters are
very helpful to make a measurement of (\ref{gbuc:acp}) feasible.
First, the $B_d$ lifetime (about $1.5ps$) is relatively large due to
the smallness of $V_{cb}\approx 0.04$, which is crucial for being
able to resolve the time dependence. Furthermore, $\Delta M$ is
sizable due to the large top quark mass such that $\Delta M$ turns out
to be of almost the same size as $\Gamma$ (see Table \ref{gbuc:t:kbb}),
which is almost perfect for optimizing the effect of mixing.
\end{itemize}

\subsection{Other Possibilities}

A case very similar to $B_d(\bar B_d)\to J/\Psi K_S$ is the CP
asymmetry for $B_d(\bar B_d)\to\pi^+\pi^-$. Here the dominant
contribution to the amplitude has CKM factor $V^*_{ub}V_{ud}$
($V_{ub}V^*_{ud}$) and consequently the relative phase between the 
mixed decay $B\to \bar B\to f$ and the direct decay $B\to f$ is given by
$V^*_{tb}V_{td}V_{ub}V^*_{ud}/(V_{tb}V^*_{td}V^*_{ub}V_{ud})=
\exp(-2i(\beta+\gamma))=\exp(2i\alpha)$.
Accordingly the CP asymmetry is a measure of $\sin 2\alpha$. The 
situation is, however, somewhat complicated by the second, 
non-negligible contribution to the decay amplitude from penguin graphs.
This contribution comes with CKM factor $V^*_{tb}V_{td}$, which, unlike
the case of $B\to J/\Psi K_S$, has a different phase than the leading
contribution ($\sim V^*_{ub}V_{ud}$). Consequently, the amplitude no
longer has the simple structure of the $B\to J/\Psi K_S$ amplitude
with its single weak phase where all hadronic uncertainties cancel,
and some poorly calculable hadronic dynamics will invariably enter the
CP asymmetry ${\cal A}_{CP}(B_d\to\pi^+\pi^-)$ (`penguin pollution').
Strategies have been devised to eliminate this uncertainty, for instance
using additional information from related modes as
$B_d(\bar B_d)\to\pi^0\pi^0$ and $B^\pm\to\pi^\pm\pi^0$ together with
isospin symmetry \cite{gbuc:GL}.
Assuming this has been achieved, $B\to\pi\pi$ determines $\sin 2\alpha$,
an example of mixing-induced CP violation just as the case of
$B\to J/\Psi K_S$ and $\sin 2\beta$. As explained before, each of these
cases considered separately represents indirect CP violation. However
any deviation from the equality $\sin 2\beta=-\sin 2\alpha$ would reveal
a direct CP violation effect \cite{gbuc:WW}
(the minus sign appears here due to the
opposite CP parities of $J/\Psi K_S$ (CP odd) and $\pi^+\pi^-$
(CP even)).

A good example of direct CP violation is provided by the decays
$B^\pm\to D^0_{(CP+)}K^\pm$. No flavor-tagging or time-dependent
measurements are required here and the asymmetries can be used to
extract the angle $\gamma$ in a clean way \cite{gbuc:GW,gbuc:ADS}.

Also $B_s$ mesons offer opportunities for interesting CP violation studies,
although they are more challenging experimentally because of the very
large oscillation frequency $\Delta M/\Gamma > 10$.
For instance, $B_s\to J/\Psi\phi$ is the $B_s$ analog of
$B_d\to J/\Psi K_S$ decay. The asymmetry is Cabibbo suppressed in this
case but would allow, in principle, a clean determination of $\eta$.
A measurement of $\gamma$ is possible with $B_s\to D^+_sK^-$ 
\cite{gbuc:ADK}.

There are many more strategies and scenarios discussed in the literature.
In our brief 
account we have focused on those cases that can yield insight into the
mechanisms of CP violation with exceptionally small theoretical
uncertainties. For general reviews see e.g. \cite{gbuc:NQ,gbuc:BF}.

\section{Conclusions}

The violation of CP symmetry has so far been observed in just five
decay modes of the long-lived neutral kaon, namely
$K_L\to\pi^+\pi^-$, $\pi^0\pi^0$, $\pi e\nu$, $\pi\mu\nu$,
$\pi^+\pi^-\gamma$. All asymmetries can be described by a single
complex parameter $\varepsilon$. The question of direct CP violation
in $K\to\pi\pi$, measured by $\varepsilon'/\varepsilon$, is still open
and currently further pursued by ongoing projects.
Although our knowledge of this phenomenon is rather limited, the
established pattern of CP violation with kaons, 
$\varepsilon\sim 10^{-3}$ and 
$\varepsilon'
\;\raisebox{-.4ex}{\rlap{$\sim$}} \raisebox{.4ex}{$<$}\; 10^{-6}$,
is well accounted for by the three generation Standard Model.
The smallness of $\varepsilon$ and $\varepsilon'$ is related to the 
size of ${\rm Im}V^*_{ts}V_{td}=A^2\lambda^5\eta\sim 10^{-4}$.
This quantity is small due to suppressed intergenerational 
quark mixing ($\sim\lambda^5$), but not due to smallness of the
CP violating phase ($\sim\eta$), which in fact is quite substantial
(typically $\eta\approx 0.3 - 0.4$).
As a consequence, large asymmetries are predicted in the $B$ meson
sector, which has many decay channels and a very rich
phenomenology.
The highlight of this latter area is
${\cal A}_{CP}(B_d\to J/\Psi K_S)\sim\sin 2\beta\approx 0.6\pm 0.2$,
exhibiting a large effect with essentially no theoretical uncertainties
and good experimental feasibility.

Theoretical progress during recent years that is of relevance for this
type of physics includes heavy quark effective theory, the calculation
of higher order QCD effects and improvements in lattice QCD
computations. In many cases a serious remaining problem is the
nonperturbative strong dynamics governing weak decay matrix elements
($\varepsilon'/\varepsilon$, $B\to\pi\pi$).
Exceptions are clean observables with practically negligible theoretical
error. The prime examples of this class are
${\cal A}_{CP}(B_d\to J/\Psi K_S)$ ($\sim\sin 2\beta$) and
$B(K_L\to\pi^0\nu\bar\nu)$ ($\sim\eta^2$).
Also important for a further understanding of CP violation is the study
of flavor-changing neutral current rare decays with small theoretical
ambiguities such as $K^+\to\pi^+\nu\bar\nu$, $B\to X_s\gamma$,
$B\to X_s e^+e^-$, $B\to l^+l^-$, $B\to X_s\nu\bar\nu$ or
$\Delta M_{B_s}/\Delta M_{B_d}$. Since the number of clean processes
is very limited, and much complementary information is needed,
all of them should be pursued as far as possible.

CP violation is intimately connected with flavor dynamics, the least
understood sector of our current Standard Model. It is therefore also
closely related to the question of electroweak symmetry breaking,
presently one of the most urgent open problems in fundamental physics.
The coming years hold great promise for decisive progress in our
knowledge about CP violation and for obtaining a clearer picture of
what may still lie behind this remarkable phenomenon.

\end{document}